\documentclass[12pt]{article}
\usepackage{graphicx}

\usepackage{amsfonts,epsfig,eurosym,color, amsmath, amsthm}
\usepackage{dsfont}
\usepackage[latin1]{inputenc}
\usepackage[T1]{fontenc}
\usepackage{pdfsync}
\usepackage{natbib}
\usepackage{rotating}
\usepackage[normalem]{ulem}
\usepackage{bbm}
\usepackage{url}

\DeclareMathOperator{\nach}{ne}

\DeclareMathOperator{\T}{^{\mathsf T}}
\def\V{\mathcal{V}}
\def\E{\mathcal{E}}
\def\G{\mathcal{G}}

\newcommand{\indep}{\mathrel{{\perp}\hspace*{-0.6em}{\perp}}}
\newcommand{\given}{\mathrel{|}}
\newcommand{\condindep}[3]{#1 \indep #2 \given #3}
\newtheorem{definition}{Definition}

\newcommand{\blind}{1}

\begin{document}

\def\spacingset#1{\renewcommand{\baselinestretch}%
{#1}\small\normalsize} \spacingset{1}


\if1\blind
{
  \title{\bf Graphical modelling of multivariate spatial point processes with continuous marks}
  \author{Matthias Eckardt\thanks{}\hspace{.2cm}\\
    Department of Computer Science, Humboldt Universit\"{a}t zu Berlin, Germany\\
    and \\
   Jorge Mateu\\
    Department of Mathematics, University Jaume I, Castell\'{o}n, Spain}
  \maketitle
} \fi

\if0\blind
{
  \bigskip
  \bigskip
  \bigskip
  \begin{center}
    {\LARGE\bf Graphical modelling of multivariate spatial point processes with continuous marks}
\end{center}
  \medskip
} \fi

\bigskip
\begin{abstract}
This paper is the second in a series of papers which combine graphical modelling and marked spatial point patterns. Extending the previous results of \cite {Eckardt2016a}, we introduce a marked spatial dependence graph model which depicts the global dependence structure of quantitatively marked multi-type points that occur in space based on the marked conditional partial spectral coherence. Most beneficial, no structural assumption with respect to the characteristics in the data are to be made prior to analysis. This approach presents a computationally efficient method of pattern recognition in highly structured and high dimensional multi-type spatial point processes where also quantitative marks are available. Unlike all previous methods, our new model permits the simultaneous analysis of all multivariate conditional interrelations. The new technique is illustrated analysing the diameter at breast hight of $37$ different tree species recorded at $10053$ locations in Duke Forest.     
\end{abstract}

\noindent%
{\it Keywords:} Continuous marks; Duke Forest; Graphical model; Multivariate spatial point pattern
\vfill

\newpage
\spacingset{1.45} 

\section{Introduction}

This paper considers the simultaneous analysis of spatial point processes where to each point (location) additional discrete-valued and real-valued information, called qualitative and quantitative marks, can be assigned. In general, our approach covers situations where only one qualitative mark and at least one, but possibly multiple, quantitative marks are present. For simplicity, this paper only focus on one qualitative and one quantitative mark, although the approach can easily be extended to multiple quantitative marks.

Although several authors have contributed to the field of marked spatial processes, the analysis of such data still remains challenging and further methodological investigations are needed. Here, both points and marks  possibly exhibit a spatial structure and various interrelations between mark-mark, point-mark or point-point might be present. Different treatments of  marks and various methodological approaches are described in  \cite{Penttinen1992}, \cite{StoyanStoyan1994}, \cite{Stoyan1995}, \cite{Mateu2000} ,  \cite{Stoyan2000}, \cite{Schlather2001}, \cite{Schlather2004},  \cite{Guan2006}, \cite{Guan2007a},  \cite{Myllymaeki2009} and \cite{MollerGhorbaniRubak2016}  among others. For a comprehensive treatment of various models and statistics for marks and points we refer the interested reader to \cite{Illian2008}, \cite{Baddeley2010} and \cite{Diggle2013}  while \cite{Baddeley2015} intensively covers the computational analysis of such data.

As qualitatively marked spatial point processes, such as species of trees or types of crops, have also been denoted as multivariate or multi-type spatial point process, we address any process with purely qualitative marks as multivariate point processes. Multivariate spatial point patterns have been analysed in various disciplines including epidemiology \citep{Diggle2005}, ecology \citep{Illian2007} or forestry \citep{Grabarnik2009, Wiegand2007}.

Different to the analysis of multivariate spatial point processes this paper additionally considers  quantitative marks and we denote any such process as multivariate marked spatial point process (MMSPP). More formally, we aim to model the global structural interrelations of a $d$-variate marked spatial point process $\boldsymbol{\Psi}=(\Psi_1,\ldots,\Psi_d)$ on $\mathds{R}^2\times\mathcal{M}$ where each $\Psi_d$ can be written as bivariate sequence of a spatial point process $Z_d\in\mathds{R}^2$ and a mark $\zeta_d=\zeta(Z_d)$ which belongs to some mark space $\mathcal{M}$ which might be Polish. For specificity we focus on data on forest stands of  $37$ different types of trees recorded in Duke Forest in North Carolina and the diameter at breast height (DBH) value measured in 2014 as a continuous mark.

Our idea is to identify the global dependence structure of a MMSPP by means of an undirected graphical model  which we term marked spatial dependence graph model (mSDGM) where the edge set graphically displays Markov properties and is related to the marked spectral coherence. Precisely, the mSDGM visualises the global conditional interrelation structure between the component processes of possibly highly complex as well as high dimensional MMSPP by using an undirected graph. Thus, the mSDGM extends the spatial dependence graph model (SDGM) as recently proposed by \cite{Eckardt2016a} for multivariate spatial point patterns to more complex data. The idea to use graphical models in the context of quantitative marks is new.

Most generally, undirected graphs have been used to capture spatial neighbourhood relations of lattice data. In contrast, the usage of undirected graphs in the context of spatial point patterns remains limited.  Apart from the SGDM \citep{Eckardt2016a}, undirected graphs denoted as neighbour networks have been discussed by \cite{Marchette2004}, \cite{Penrose2003,Penrose2005} and \cite{Penrose2001}. These graphs are random graph models where single events are represented by distinct nodes. Alternatively, undirected graphs have been introduced as a new spatial domain within the class of point processes on linear networks by \cite{Okabe2001}, \cite{Ang2010}, \cite{Ang2012} and \cite{Baddeley2014} and also, besides other graph structures, by \cite{Eckardt2016b} and \cite{Eckardt2016c}.  

This remainder of this paper is organized as follows: Section \ref{sec:1} briefly introduces the fundamental graph theoretical concepts, describe the properties of spatial point processes and marked spatial point processes in the spatial and also the frequency domain and define the mSDGM. An application of the mSDGM to data on forest stands is given in Section \ref{sec:2}.  Finally, the theoretical results are discussed in Section \ref{sec:3}.  

\section{Graphical modelling of spatial point patterns}\label{sec:1}

We now extend the class of spatial dependence graph models with respect to the exploration of global conditional interrelations which are present in a MMSPP. Here we assume that the qualitative mark contains at least $3$ disjoint types. This new graphical model is linked to the marked conditional spectral properties of a finite set of randomly occurring marked points of different types in a bounded region. Hence, we relate the dependence structure of a multivariate marked point process to the adjacency structure encoded in a graph. To discuss the mSDGM in detail, we first need to introduce the basic notation and terminology of graph theory. For a rigorous treatment of graph theory we refer the interested reader to \citet{Bondy2008} and also \citet{Diestel2010}. Formally, a graph consists of a pair $\G=(\V,\E)$ where $\V=\left\{v_1,\dots,v_k\right\}$ is a finite set of vertices or nodes, and $\E\subseteq\V\times\V$ is a finite set of edges joining the vertices, where $\E(\G)\cap\V(\G)=\emptyset$. Depending on the shape and type of the edges, we can define several different graphs although we only consider undirected graphs. Precisely, we only allow for undirected edges joining pairs of vertices. Any pair of vertices which is joined by an edge is called adjacent and the set of all adjacent nodes of a distinct node $v_j$ is the neighbourhood $\nach\left(v_j\right)=\lbrace
v_i: (v_i,v_j)\in\E(\G)\rbrace$. A walk is a sequence of potentially repeating vertices and edges in $\G$. A special case of a walk is a path which passes through every node of a sequence exactly once. If every distinct pair of vertices in $\G$ is joined by a path, $\G$ is said to be connected. A component is a non-empty maximal connected subgraph $\G'$ of $\G$ such that every distinct pair of nodes is joined by a path in $\G'$. Finally, a set $\mathcal{S}$ is called a separating set or ij-vertex-cut if for any partition $v_i, v_j\in\V(G)$ and $\mathcal{S}\subset\V(\G)\backslash\lbrace v_i,v_j\rbrace$  of the node set, $v_i$ and $v_j$ are not in the same component in $\G\setminus\mathcal{S}$.

Graphical models now link graph theory and probability theory such that conditional independence statements can be read of missing edges in the graph. For a profound treatment of different graphical models, the interested reader is referred to  \cite{Pearl1988},  \cite{Cox1996}, \cite{Lauritzen1996}, \cite{Cowell1999}, \cite{Edwards2000}, \cite{Spirtes2000}, \cite{Whittaker2008} and \cite{Koller2010}.

Although these models have originally been related to cross-sectional data, numerous dynamic graphical models have recently been derived in the temporal and, less often, the frequency domain. Time domain models for point process data include the local dependence graph model \citep{Didelez2000, Didelez2007, Didelez2008}, the dynamic path analysis model  \citep{Fosen2006,Aalen2008,Martinussen2010} as well as the graphical duration model \citep{Dreassi2007,Gottard2007,Gottard2007a} besides others. Similar models for the frequency domain have been presented by  \citet{Brillinger1996}, \cite{Dahlhausetal1997} and also \citet{Eichler2003}. A comprehensive review of temporal graphical models is given in \citet{Eckardt2016}. Similar to \cite {Eckardt2016a}, the mSDGM extends these frequency domain models for multivariate marked spatial point patterns.  

\subsection{Properties of spatial point processes}\label{spprob}

To discuss the spectral analysis and the mSDGM in detail, we first need to discuss first- and second-order properties of unmarked spatial point processes for the spatial domain. First-order properties are related to the mean number of events per unit area while second-order properties are related to the covariance between the number of points in two distinct regions.

Generally, we denote the location of a randomly occurring point within a bounded region $\mathbf{S}\subset\mathds{R}^2$ as  $\mathbf{s}=(x, y)$. For the $i$-th type of point let $N_i(\mathbf{s})$ denotes the number of observed events at location $\mathbf{s}$ and $dN_i(\mathbf{s}) = N_i(\mathbf{s}+d\mathbf{s}) - N_i(\mathbf{s})$ expresses the number of observed events of type $i$ within a infinitesimal region containing $\mathbf{s}$. An in-depth discussion of the statistical analysis of spatial point processes is given in \citet{Diggle2002}, \citet{Moller2004} and \cite{Illian2008}.  

Usually, first-order properties of a spatial point process are expressed by means of the first-order intensity function. Following the notation of \citet{Diggle2002, Diggle2013}, the first-order intensity function is given as 
\[
\lambda_i(\mathbf{s})=\lim_{|d\mathbf{s}|\rightarrow 0}\left\{\frac{\mathds{E}\left[ N_i(d\mathbf{s}))\right]}{|d\mathbf{s}|}\right\}, \mathbf{s}\in\mathbf{S}.
\]  
where $|d\mathbf{s}|$ denotes the area covered by $d\mathbf{s}$. 

For the second-order properties of a spatial point pattern, one possibility is to use the second-order intensity function $\lambda_{ii}(\mathbf{s,s'})$ which is also connected to the highly prominent reduced second-order moment function also known as Ripleys' $K$-function \citep{Ripley1976}.
Formally, for a pair of locations $\mathbf{s}=(x,y)$ and $\mathbf{s'}=(x',y')$ we have
\[
\lambda_{ii}(\mathbf{s,s'})=\lim_{|d\mathbf{s}|,|d\mathbf{s}|\rightarrow 0}\left\{\frac{\mathds{E}\left[N_i(d\mathbf{s})N_i(d\mathbf{s'})\right]}{|d\mathbf{s}||d\mathbf{s'|}}\right\}, \mathbf{s}\neq\mathbf{s'}, \mathbf{s},\mathbf{s'}\in\mathbf{S}.
\]

Alternatively, we can describe the second-order properties using the covariance density function. Different from the second-order intensity function, this function efficiently describes the theoretical properties of spatial point patterns. Focussing on  multivariate spatial point patterns, a further distinction in the auto-covariance and the cross-covariance density function is possible. These functions encode the component-specific within and between variation. The auto-covariance density function is defined as
\begin{equation}
\gamma_{ii}(\mathbf{s,s'})=\lim_{|d\mathbf{s}|,|d\mathbf{s'}|\rightarrow 0}\left\{\frac{\mathds{E}\left[\lbrace N_i(d\mathbf{s})-\lambda_i(d\mathbf{s})\rbrace \lbrace N_i(d\mathbf{s'})-\lambda_i(d\mathbf{s'})\rbrace\right]}{|d\mathbf{s}||d\mathbf{s'|}}\right\}
\label{autocov}
\end{equation}
and can be obtained from the second-order intensity function as
\[
\gamma_{ii}(\mathbf{s,s'})=\lambda_{ii}(\mathbf{s,s'})-\lambda_i(\mathbf{s})\lambda_i(\mathbf{s'}).
\]
Similarly, we obtain the cross-covariance density function for any two disjoint events $i$ and $j$ as
\begin{equation}
\gamma_{ij}(\mathbf{s,s'})=\lim_{|d\mathbf{s}|,|d\mathbf{s'}|\rightarrow 0}\left\{\frac{\mathds{E}\left[\lbrace N_i(d\mathbf{s})-\lambda_i(d\mathbf{s})\rbrace\lbrace N_j(d\mathbf{s'})-\lambda_j(d\mathbf{s'})\rbrace\right]}{|d\mathbf{s}||d\mathbf{s'|}}\right\}.
\label{crosscov}
\end{equation}

For orderly processes, which imply that only one event can occur at a particular location,  \eqref{autocov} and \eqref{crosscov} include the case when $\mathbf{s}=\mathbf{s'}$. Precisely, for orderly processes we have $
\mathds{E}\left[\lbrace N_i(d\mathbf{s})\rbrace^2\right]=\lambda_i(\mathbf{s})|d\mathbf{s}|$. The integration of this expression into the covariance density function leads to Bartlett's complete auto-covariance density function $\kappa_{ii}(\cdot)$  \citep{Bartlett1964}, namely
\begin{equation}\label{completecov}
\kappa_{ii}(\mathbf{s,s'})=\lambda_i(\mathbf{s})\delta(\mathbf{s}-\mathbf{s'})+\gamma_{ii}(\mathbf{s,s'})
\end{equation}
where $\delta(\cdot)$ denotes a two-dimensional Dirac delta function.

As before, dealing with multivariate processes we can also distinguish between the complete auto-covariance and the complete cross-covariance density function. Following \cite{Mugglestone1996a}, we then have for the complete cross-covariance for events of types $i$ and $j$ that $\kappa_{ij}(\mathbf{s,s'})=\gamma_{ij}(\mathbf{s,s'})$ and $\kappa_{ji}(\mathbf{s,s'})=\gamma_{ji}(\mathbf{s,s'})$.

Based on the previous results, we now consider the case of marked spatial point processes. A well-known correlation-based characteristic of a marked spatial point process is the mean product of marks sited at distance $r$ apart and will be denoted by $U(r)$. For a stationary and isotropic process, such that the probabilistic statements about a process are invariant under translation and also invariant under rotation, $U(r)$ follows as
\begin{equation}\label{eq:meanmarks}
U(r)=\lambda^2g(r)k(r)da_1da_2
\end{equation}
where $g(r)$ is the pair correlation function, $k(r)$  is  the mark correlation function and $da_1$ and $da_2$
are two infinitesimal small areas separated by a  distance $r$. Obviously, for an unmarked point process we have that $k(r)=1$ and for a complete spatial random process we have that $g(r)=1$.
An extension of \eqref{eq:meanmarks} for anisotropic processes has been presented by \cite{StoyanStoyan1994} replacing the distance $r$ by the polar form $(r,\vartheta)$ yielding 
\begin{equation}\label{eq:meanmarksStoyan}
U(r,\vartheta)=\lambda^2g(r,\vartheta)k(r,\vartheta)da_1da_2.
\end{equation}
Another modification of \eqref{eq:meanmarks} was proposed by \cite{Capobianco1998} considering a Cartesian framework for point process data observed over rectangular regions, where the polar form of \eqref{eq:meanmarksStoyan} is replaced by a Cartesian distance using a city-block metric.

\subsection{Spectral properties of spatial point processes}\label{sec:spectralproperties}

This section introduces spectral properties of spatial point processes. As before, we first discuss the unmarked case and then generalise the results for marked spatial point processes. 

Generally, Fourier transformations and spectral analysis techniques determine the presence of periodic structures in spatial point processes and present a complementary approach to distance-related methods. Most beneficially, in contrast to inter-distance techniques and statistical models in the spatial domain, frequency domain methods do not require any prior distributional assumptions and allow for anisotropic or non-stationary processes, and also for different scales.

Although spectral techniques have become a well-known and frequently applied method in the periodic analysis of time series data, their application with respect to spatial point processes remain very limited and only a few methodological and applied contributions exist. Even less work has been published for marked spatial point pattern. The periodic analysis of temporal point processes by means of spectral techniques has been pioneered by  \citet{Bartlett1963} and later by \citet{Brillinger1972}. The first extension to two-dimensional point processes was presented in the seminal paper by \citet{Bartlett1964}. A profound treatment of spectral properties with respect to spatial point processes is given in \cite{Renshaw1983},  \citet{Renshaw1984}, \citet{Renshaw1997} and \citet{Mugglestone1996a}, \citet{Mugglestone1996b},  \cite{Mugglestone2001}. The spectral analysis of marked spatial point processes has first been covered in
\cite{Renshaw2002}. In addition, \cite{Saura2006} discuss the estimation of mark functions by means of spectral techniques.   

Generally, in order to discuss the spectral properties theoretically, we assume the spatial point process to be orderly - such that multiple coincident events can not occur - and second-order stationary. Second-order stationarity implies that the first-order intensity function $\lambda_i(\mathbf{s}), \mathbf{s}\in\mathbf{S}$ is constant over a finite region $\mathbf{S}\subset \mathds{R}^2$ while the covariance density function $\gamma_{ij}(\mathbf{s,s'})$ depends on $\mathbf{s}$ and $\mathbf{s'}$ only through $\mathbf{c}=\mathbf{s}-\mathbf{s'}$. For a $d$-variate process the notion of stationarity implies that all $d$ processes are marginally and jointly stationary. Consequently, we have $\gamma_{ii}(\mathbf{s,s'})=\gamma_{ii}(\mathbf{c})$ and also $\kappa_{ii}(\mathbf{s,s'})=\kappa_{ii}(\mathbf{c})$. For the covariance density function we notice that $\gamma_{ij}(\mathbf{s,s'})=\gamma_{ji}(\mathbf{s',s})$ such that stationarity also implies that $\gamma_{ij}(\mathbf{c})=\gamma_{ji}(-\mathbf{c})$ and $\kappa_{ij}(\mathbf{c})=\kappa_{ji}(-\mathbf{c})$ (cf. \citet{Mugglestone1996a, Mugglestone1996b}).   

For a second-order stationary spatial point process, the auto-spectral density function for an event $i$ at frequencies $\boldsymbol{\omega}=(\omega_1,\omega_2)$ appears as the Fourier transform of the complete auto-covariance density function of $N_i$,  
 \begin{equation}\label{fouriereq}
\begin{split}
f_{ii}(\boldsymbol{\omega}) &= \int \kappa_{ii}(\mathbf{c})\exp(-\iota\boldsymbol{\omega}^{\T}\mathbf{c})d\mathbf{c}\\
&=\int^\infty_{-\infty}\int^\infty_{-\infty}\kappa_{ii}(c_1,c_2)\exp\lbrace-\iota(\omega_1c_1+\omega_2c_2)\rbrace dc_1dc_2
\end{split}
\end{equation}
where $\iota=\sqrt{-1}$ and $\boldsymbol{\omega}^{\T}$ denotes the transpose of $\boldsymbol{\omega}$. As described in \citet{Brillinger1981} and \cite{Brockwell2006} with respect to time series, the auto-spectrum can be understood as the decomposition of $\kappa_{ii}$ into a periodic function of frequencies $\boldsymbol{\omega}$.

From expression \eqref{fouriereq}, the complete auto-covariance density function can uniquely be recovered via inverse Fourier transformation,
\begin{equation}\label{inversekappa}
\kappa_{ii}(\mathbf{c}) = \int f_{ii}(\boldsymbol{\omega})
\exp\left(\iota\boldsymbol{\omega}^{\T}\mathbf{c}\right)d\boldsymbol{\omega}.
\end{equation}

Substituting for $\kappa_{ii}(\mathbf{c})$ from \eqref{completecov} finally leads to 
 \begin{equation}\label{fouriereqfinal}
f_{ii}(\boldsymbol{\omega}) =\lambda_i+\int^\infty_{-\infty}\int^\infty_{-\infty}\gamma_{ii}(c_1,c_2)\exp\lbrace-\iota(\omega_1c_1+\omega_2c_2)\rbrace dc_1dc_2.
\end{equation}

Similarly, the cross-spectral density function is given as the Fourier transform of the complete cross-covariance density function,
\begin{equation}\label{crossspectrakappa}
f_{ij}(\boldsymbol{\omega}) = \int \kappa_{ij}(\mathbf{c})\exp(-\iota\boldsymbol{\omega}^{\T}\mathbf{c})d\mathbf{c},
\end{equation}
which measures the linear interrelation of components $N_i$ and $N_j$. Thus, two processes are said to be uncorrelated at all spatial lags if and only if the corresponding spectrum is zero at all frequencies. In addition, since $\kappa_{ij}(\mathbf{c})=\kappa_{ji}(-\mathbf{c})$ we equivalently have that also $f_{ij}(\mathbf{c})=f_{ji}(-\mathbf{c})$ and thus it is sufficient to calculate only one cross-spectrum (cf. \citet{Bartlett1964},  \cite{Mugglestone1996a,Mugglestone1996b}). 

We now discuss the Fourier transformation of marked spatial point patterns. Adopting the results of \cite{Renshaw2002}, we obtain the marked auto- and cross-spectral density function by replacing the complete auto- and cross-covariance function by the a suitable version of $U(\cdot)$ such as introduced in \eqref{eq:meanmarks} or \eqref{eq:meanmarksStoyan}. Here, similar to the $\kappa(\cdot)$, we express the mean product of marks for the $i$-th multivariate marked process by $U_{ii}(\cdot)$ and between the $i$-th and the $j$-th multivariate marked process by  $U_{ij}(\cdot)$.

Usually, the cross-covariance function could be asymmetric, namely $\gamma_{ij}(\mathbf{c})\neq \gamma_{ji}(\mathbf{-c})$, such that the cross-spectrum is a complex-valued function. As discussed in \citet{Priestley1981} and \citet{Chatfield1989}, a common procedure in time series analysis is to split the complex-valued cross-spectrum into the real and the imaginary part, namely into the co-spectrum $C_{ij}(\boldsymbol{\omega})$ and quadrature spectrum $Q_{ij}(\boldsymbol{\omega})$ at frequencies $\boldsymbol{\omega}$. Thus, the marked cross-spectrum can be decomposed in terms of Cartesian coordinates as
\[
\begin{split}
f_{ij}(\boldsymbol{\omega})&=\frac{1}{2\pi}\sum^\infty_{\mathbf{c}=-\infty}\cos(\boldsymbol{\omega}^{\T}\mathbf{c})U_{ij}(\cdot)-\iota \frac{1}{2\pi}\sum^\infty_{\mathbf{c}=-\infty}\sin(\boldsymbol{\omega}^{\T}\mathbf{c})U_{ij}(\cdot)\\
&=C_{ij}(\boldsymbol{\omega})-\iota Q_{ij}(\boldsymbol{\omega}).
\end{split}
\]

Another possibility is to express the marked cross-spectrum by means of polar coordinates which will not be further investigated here.  For a detailed discussion of the spectral properties of multivariate spatial point processes we refer the interested reader to \cite{Eckardt2016a}.

Although the marked cross-spectrum expresses the linear interrelation between two component processes, it is often preferable to use the marked spectral coherence as a rescaled version of the marked cross-spectrum. The marked spectral coherence is then defined as
\begin{equation}\label{eq:Coh}
\vert R_{ij}(\boldsymbol{\omega})\vert^2=\frac{f_{ij}(\boldsymbol{\omega})^2}{\left[f_{ii}(\boldsymbol{\omega})f_{jj}(\boldsymbol{\omega})\right]}
\end{equation}
and measures the linear relation of two components. Different from the marked auto-spectrum, resp. marked cross-spectrum, we have that $0\leq \vert R_{ij}(\boldsymbol{\omega})\vert^2\leq 1$.

However, the marked spectral coherence is not able to distinguish between direct and induced interrelations. In order to control for the linear effect of all remaining component processes $\psi_{V\backslash\lbrace i,j\rbrace}$ on pairwise linear interrelations between $\psi_i$ and $\psi_j$, we adopt the framework of partialisation. Thus, in analogy with graphical modelling of multivariate data, we are interested in the linear interrelation between $\psi_i$ and $\psi_j$ that remains after removal of the linear effect of all alternative component processes. In this respect, the partial cross-spectrum $f_{ij\given\V\backslash\lbrace i,j\rbrace}(\boldsymbol{\omega})$ follows as the cross-spectrum of the residual processes $\epsilon_i$ and $\epsilon_j$ which results from the elimination of the linear effect of $\psi_{V\backslash\lbrace i,j\rbrace}$ on $\psi_i$ and $\psi_j$. So, we have $f_{ij\given\V\backslash\lbrace i,j\rbrace}(\boldsymbol{\omega})=f_{\epsilon_i\epsilon_j}(\boldsymbol{\omega}).$

As a well-known result, one can compute the partial marked cross-spectrum by applying \citet[Theorem 8.3.1.]{Brillinger1981} using the formula
\begin{equation}\label{partial.formula}
f_{ij\given\V\backslash\lbrace i,j\rbrace}(\boldsymbol{\omega})=f_{ij}(\boldsymbol{\omega})-f_{i\V\backslash\lbrace i,j\rbrace}(\boldsymbol{\omega})f_{\V\backslash\lbrace i,j\rbrace\V\backslash\lbrace i,j\rbrace}(\boldsymbol{\omega})^{-1}f_{\V\backslash\lbrace i,j\rbrace j}(\boldsymbol{\omega})
\end{equation}
where \[
f_{i\V\backslash\lbrace i,j\rbrace}(\boldsymbol{\omega})=\left[f_{i1}(\boldsymbol{\omega}), \ldots,f_{ii-1}(\boldsymbol{\omega}), f_{ii+1}(\boldsymbol{\omega}),\ldots,f_{ij-1}(\boldsymbol{\omega}), f_{ij+1}(\boldsymbol{\omega}), \ldots, f_{ik}(\boldsymbol{\omega}) \right].
\]
An alternative solution was presented by \cite{Dahlhaus2000} which is less computerintensive. For a detailed discussion of alternative calculations of partial spectral densities we again refer to \cite{Eckardt2016a}. 

As before, we can now compute the partial marked spectral coherences by  rescaling of the partial marked cross-spectrum, 
\begin{equation}
|R_{ij\given\V\backslash\lbrace i,j\rbrace}(\boldsymbol{\omega})|^2=\frac{f_{ij\given\V\backslash\lbrace i,j\rbrace}(\boldsymbol{\omega})^2}{\left[f_{ii\given\V\backslash\lbrace i,j\rbrace}(\boldsymbol{\omega})f_{jj\given\V\backslash\lbrace i,j\rbrace}(\boldsymbol{\omega})\right]}.
\end{equation}
Equivalently, the partial marked spectral coherence also follows as $|R_{ij\given\V\backslash\lbrace i,j\rbrace}(\boldsymbol{\omega})|^2$ from $g_{ij}(\boldsymbol{\omega})$, where
\begin{equation}
\label{InverseSpectraRxyz}
R_{ij\given\V\backslash\lbrace i,j\rbrace}(\boldsymbol{\omega})=-\frac{g_{ij}(\boldsymbol{\omega})}{\left[ g_{ii}(\boldsymbol{\omega})g_{jj}(\boldsymbol{\omega})\right]^{\frac{1}{2}}}
\end{equation}
as proven in \cite{Dahlhaus2000}.
Under regularity assumptions, we can also define the absolute rescaled marked inverse as
\begin{equation}
\vert d_{ij}(\boldsymbol{\omega})\vert=\frac{\vert g_{ij}(\boldsymbol{\omega}) \vert}{\left[g_{ii}(\boldsymbol{\omega})g_{jj}(\boldsymbol{\omega})\right]^{\frac{1}{2}}}
\end{equation}
which measures the strength of the linear partial interrelation between $\psi_i$ and $\psi_j$ at frequencies $\boldsymbol{\omega}$. As shown in \cite{Dahlhaus2000}, we then have
\begin{equation}
d_{ij}(\boldsymbol{\omega})=-R_{ij\given\V\backslash{\lbrace i,j\rbrace}}(\boldsymbol{\omega})
\end{equation} 
such that we can obtain the partial spectral coherence from the negative of the absolute rescaled inverse.

\subsection{Spatial dependence graph model}

This section defines the marked spatial dependence graph model. Here, the idea is to express the partial interrelation structure of a marked multivariate spatial point pattern by means of an undirected graph. 

Given an observed multivariate marked spatial point pattern $\boldsymbol{\psi}_\V$ we identify the nodes of an undirected graph with the components of a multivariate spatial counting process. For any $\psi_i$ component processes $\psi_j$ of $\boldsymbol{\psi}$ which are conditional orthogonal after elimination of all remaining components $\psi_{\V\setminus\lbrace i,j\rbrace}$ we have that the two vertices $v_i$ and $v_j$ are unconnected in the undirected graph. As previously discussed, $\psi_i$ and $\psi_j$ are said to be conditional orthogonal if and only if the partial marked spectral coherence vanishes at all frequencies $\boldsymbol{\omega}$. This holds whenever the partial marked cross-spectrum $f_{ij\given\V\backslash\lbrace i,j\rbrace}(\boldsymbol{\omega})$, the inverse $g_{ij}(\boldsymbol{\omega})$ or equivalently the absolute rescaled inverse $d_{ij}(\boldsymbol{\omega})$ is zero at all frequencies $\boldsymbol{\omega}$.  From this, we define a mSDGM as follows.
\begin{definition}
Let $\boldsymbol(\psi)_\V$ be a multivariate marked spatial point pattern. A marked spatial dependence graph model is an undirected graphical model $\G=(\V,\E)$ in which any $v_i\in\V(\G)$ encodes a component of $\boldsymbol{\psi}_\V$ and $\E(\G)=\lbrace (v_i,v_j): R_{i,j\given \V\backslash\lbrace i,j\rbrace}(\boldsymbol{\omega})\neq 0\rbrace$ such that
\begin{equation*}
\condindep{\lbrace \psi_i\rbrace}{\lbrace \psi_j\rbrace}{\lbrace \psi_{\V\backslash\lbrace i,j\rbrace}\rbrace}\Leftrightarrow (v_i,v_j)\notin\E(\G).
\end{equation*}
\end{definition}

So, a mSDGM encodes conditional orthogonality relations between different components of a multivariate marked spatial point process by means of an undirected graph, Precisely, these conditional orthogonality relations are expressed by missing edges. 

Besides, several additional relations can be read from the graph structure as the statement $\condindep{\lbrace \psi_i\rbrace}{\lbrace \psi_j\rbrace}{\lbrace \psi_{\V\backslash\lbrace i,j\rbrace}\rbrace}$ additionally imposes that $\lbrace \psi_{\V\backslash\lbrace i,j\rbrace}\rbrace$ is a separator which intersects all paths from $\lbrace \psi_i\rbrace$ to $\lbrace \psi_j\rbrace$. This also means that $\lbrace \psi_i\rbrace$ and $\lbrace \psi_j\rbrace$ are not in the same component in $\G\setminus \lbrace \psi_{\V\backslash\lbrace i,j\rbrace}\rbrace$.

\subsection{Estimation of spectral densities}

We now concern the estimation of the marked auto- and marked cross-periodograms. Assume we have observed a $d$-variate spatial marked point pattern within a rectangular region $\mathbf{S}\subset\mathds{R}^2$ with sides of length $l_x$ and $l_y$ and let $\lbrace\mathbf{s}_i\rbrace=\lbrace(x_i,y_i)\rbrace, i=1,\ldots,N_i$ denote the locations of points of type $i$ and $\zeta_{i}$ the corresponding real-valued mark.  Respectively, $\lbrace\mathbf{s}_j\rbrace$ are the locations of points of type $j$. Then, we obtain the marked auto- and marked cross-periodograms from a discrete Fourier transform (DFT) of the marked locations $\lbrace\mathbf{s}_i\rbrace$ and $\lbrace\mathbf{s}_j\rbrace$. Thus, the DFT for the marked locations of type $i$ is given as
\begin{align*}
F_i(p,q)&=(l_x, l_y)^{- \frac{1}{2}}\sum_{i=1}^{N_i}(\zeta_{i}-\bar{\zeta_i})\exp\left(- 2\pi\iota N_i^{-1}\left(px_i + qy_i\right)\right)\\
&= A_i(p,q)+\iota B_i(p,q)
\end{align*}
where $p=0,1,2,\ldots,~q=0,\pm 1,\pm 2,\ldots$ and $ \bar{\zeta}_{i}$ is the mean over all marks for locations of type $i$ (cf. \cite{Renshaw2002}). From this, we can compute the marked auto-periodogram for frequencies $\boldsymbol{\omega}=(2\pi p/N,2\pi q/N)$ as 
\begin{eqnarray}\label{eq:autoperiodogram}
\hat{f}_{ii}(\boldsymbol{\omega})&=&F_i(p,q)\bar{F}_i(p,q)\\
&=& \lbrace A_i(p,q)\rbrace^2+\lbrace B_i(p,q)\rbrace^2.\nonumber
\end{eqnarray}
Here, $\bar{F}_i$ denotes the complex conjugate of $F_i$. 

Again we have that $\hat{f}_{ii}(\boldsymbol{\omega})=\hat{f}_{ii}(-\boldsymbol{\omega})$ such that it suffices to compute the marked periodogram for $p = 0, 1, \ldots, 16$ and $q = - 16, \ldots, 15$ (cf. \citet{Renshaw1983, Mugglestone1996a}).

The marked cross-periodogram for frequencies $\boldsymbol{\omega}=(2\pi p/N,2\pi q/N)$ can be computed similarly such that
\begin{equation}\label{eq:crossperiodogram}
\hat{f}_{ij}(\boldsymbol{\omega})=F_i(p,q)\bar{F}_j(p,q).
\end{equation}

However, to omit bias in \eqref{eq:autoperiodogram} and \eqref{eq:crossperiodogram} at low frequencies, $\lbrace\mathbf{s}_i\rbrace$ and $\lbrace\mathbf{s}_j\rbrace$ are usually standardised or rescaled to the unit square prior to analysis (cf. \citet{Bartlett1964, Mugglestone1996a}). 
Then, assuming that the locations have been scaled to the unit square, the DFT for events of type $i$ reduces to 
\[
F_i(p,q)=\sum^{N_i}_{i=1}(\zeta_{i,l}-\bar{\zeta_i})\exp(-2\pi\iota(px_i+qy_i)).
\] 
 
\section{Application to forest stands: the Duke forest data}\label{sec:2}

This section illustrates the application of the mSDGM to forest data recorded in the Duke Forest. This forest is located in Durham, Orange and Alamance counties in North Carolina (USA), and covers an area of 7000 acres of forested land as well as open fields. The forest is owned and managed by the Duke University for research and teaching since 1931 and has  spatially been analysed by several authors including \cite{Palmer1990}, \cite{Banerjee2004}, \cite{Xi2008},  \cite{Leininger2014}, \cite{Leininger2016},  \cite{Shirota2016} and  \cite{Terres2016} among others. Most of the analysis of this data were based a subset of at most 3 botanic tree species. For example, \cite{Shirota2016} considered a multivariate log-Gaussian Cox process for a subset of 3 botanic tree species (red maple, carolina buckthorn and sweetgum). In addition to the spatial coordinates and the tree species, the diameter at breast height (DBH) was also recorded.

Different from all previous research, we preselected a sample of $10053$ locations from  $14992$ records excluding cases with missing DBH information. We note that classical first- and second-order characteristics are prohibitive time-consuming with such a number of events and with so many species. We have found no published paper with such dimension for the number of types of events. Our procedure covers the analysis of the spatial interrelation of $37$ different, quantitatively marked,  botanic tree species recorded in the Duke Forest. This species are american beech ($n=26$), american elm ($n=121$),
 american holly ($n=45$), american hornbeam ($n=171$)
 black cherry ($n=33$), black oak ($n=16$),  blackgum ($n=276$),  blackhaw ($n=26$),  carolina buckthorn ($n=921$),  common persimmon ($n=40$),  downy arrowwood ($n=24$),  eastern redcedar ($n=325$),  eastern rudbud ($n=159$),  flowering dogwood ($n=770$),  loblolly pine ($n=333$),  mockernut hickory($n=361$),  northern red oak ($n=46$),  pignut hickory ($n=281$),  possumhaw ($n=49$),  post oak ($n=34$),  red maple ($n=2437$), 
 red mulberry ($n=30$),  rusty blackhaw ($n=24$), 
 shortleaf pine ($n=43$),  sourwood ($n=96$), 
 southern red oak ($n=33$),  southern sugar maple ($n=48$),  sweetgum ($n=1507$),  tree of heaven ($n=21$),  tuliptree ($n=291$),  unspecified elm ($n=2$),  virginia pine ($n=9$),  white ash ($n=482$),  
 white fringetree ($n=13$),  white oak ($n=209$), 
 willow oak ($n=11$) and finally  winged elm ($n=740$). For each tree species, we calculated marked spectral densities based on demeaned mark values. 
 
From this, we computed the mSDGM as previously described based on the absolute rescaled marked inverse spectral density. To analyse the variation in strength of the spatial interrelations, we considered three different threshold $\alpha$ which expresses a weak ($\alpha=0.3$), an intermediate ($\alpha=0.6$) and a strong spatial structural dependence ($\alpha=0.9$). The analysis was carried out in {\texttt{R}} using the \texttt{sppgraph} package \citep{Eckardt2016d}.

\begin{figure}
\centering
\makebox{\includegraphics[scale=0.55]{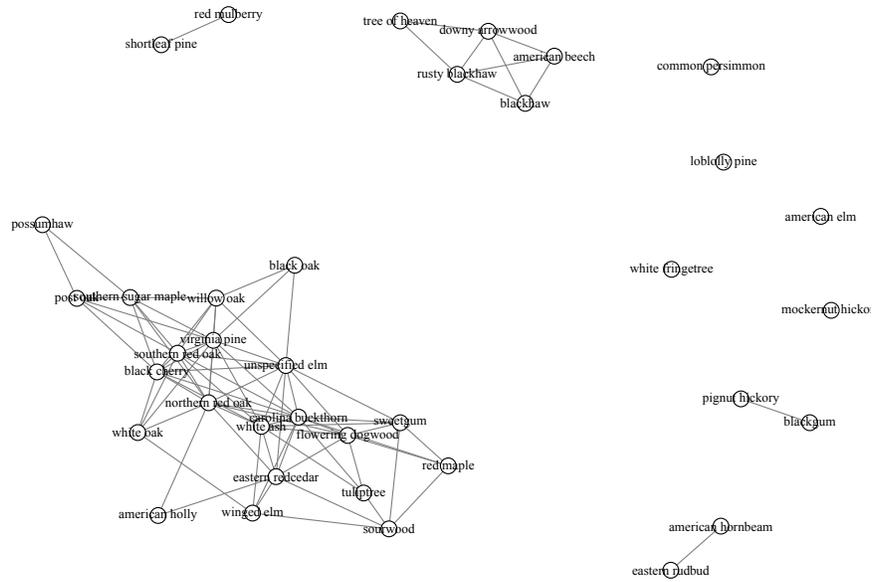}}
      \caption{\label{fig:mark30} Marked spatial dependence graph model for the $37$ different tree species, and DBH as quantitative mark for a threshold level of $\alpha=0.3$}
\end{figure}

The mSDGM for $\alpha=0.3$ is shown in Figure \ref{fig:mark30}. Here, we found $5$ isolated nodes (american elm, common persimmon, loblolly pine, white fingertree and mockernut hickory) which indicates that the spatial distribution of the DBH value of all 5 trees are independent from any other tree recorded in the analysed data. In addition, $3$ pairwise interrelations are shown as well as a $5$-node and a $21$-node subgraph. For the paired nodes we conclude that the spatial distribution of the DBH of e.g. shortleaf pine conditional on all remaining component processes only depends on the marked process of red mulberries. Similar interpretations can be made for the remaining subgraphs.    

\begin{figure}
\centering
\makebox{\includegraphics[scale=0.55]{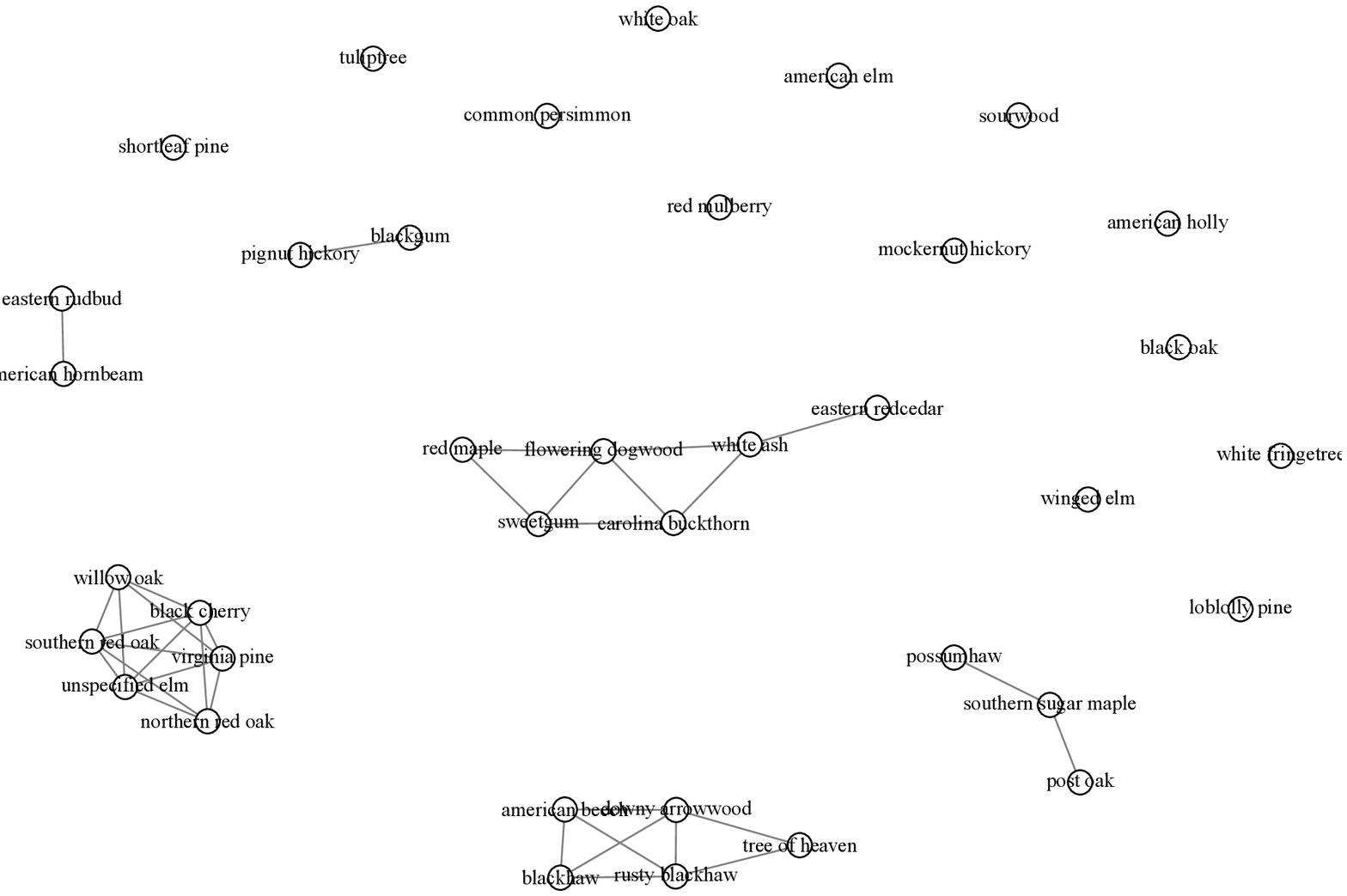}}
      \caption{\label{fig:mark60} Marked spatial dependence graph model for the $37$ different tree species, and DBH as quantitative mark for a threshold level of $\alpha=0.6$}
\end{figure}

Considering the mSDGM for $\alpha=0.6$ we observed a changing interrelation structure and an increase in isolated nodes. The corresponding graph is depicted in  
Figure \ref{fig:mark60}. For example, we observed that for the intermediate interrelation the edge between the processes pair shortleaf and red mulberries, as previously observed for $\alpha=0.3$, disappeared. In total, we now have $13$ isolated nodes, $2$ connected pairs and $4$ subgraphs with at least $3$ nodes. For this, we also found that the $5$-node subgraph of $\alpha=0.3$ remains unchanged while the original $21$-vertice subgraph splitted in $3$ subgraphs.
  
\begin{figure}
\centering
\makebox{\includegraphics[scale=0.55]{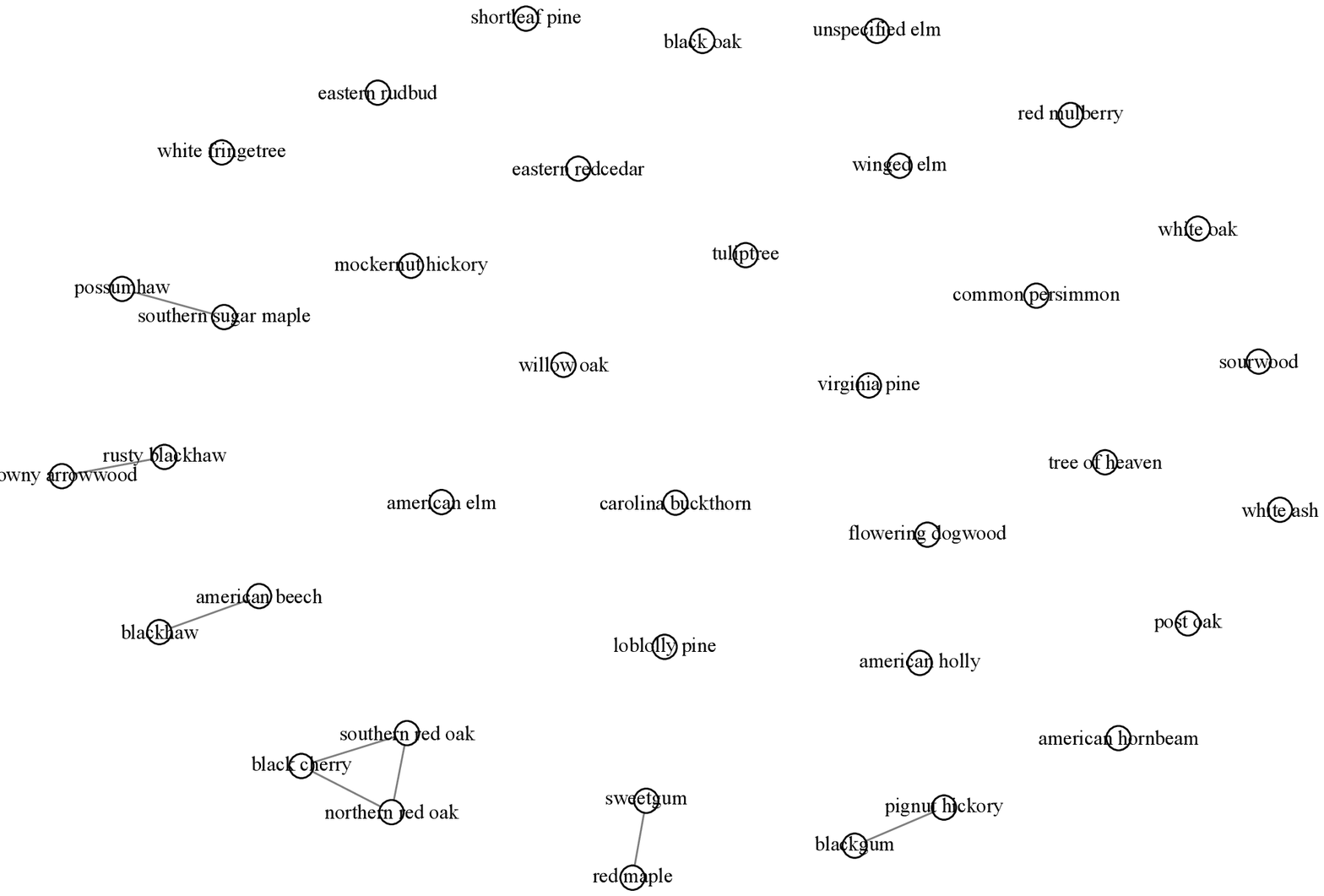}}
      \caption{\label{fig:mark90} Marked spatial dependence graph model for the $37$ different tree species, and DBH as quantitative mark for a threshold level of $\alpha=0.9$}
\end{figure}

We also note the strong interrelation structure as shown in Figure \ref{fig:mark90} for $\alpha=0.9$. Here, only one triangle ($3$-node subgraph) remained. This indicates that a strong structural dependence in the spatial distribution of the DBH values between black cherry, northern red oak and southern red oak remains after the linear effect of all remaining DBH marked tree species have been eliminated. Besides, $5$ pairwise interdepencies remain such as an edge joining sweetgum and maple. 

\section{Conclusions}\label{sec:3}
This paper has introduced a new exploratory formalism for the simultaneous exploration of global spatial dependencies for the analysis of multivariate marked spatial point patterns. The proposed model is highly computationally efficient and fast and possible applications exist in various disciplines. The mSDGM results in a comprehensive easy-to-read description of the global interrelations which exist in possibly high-dimensional spatial point patterns and is not affected by the number of observations taken into account. Thus, this paper efficiently contributes to the emerging field of high dimensional (open) data sets and the growing demand of statistical algorithms. The computation of the mSDGM is made available in \texttt{R} by \cite{Eckardt2016d} and can be applied to different data.    

The discussed linkage of graphical modelling to complex spatial point processes offers new insights into the spatial behaviour of possibly interdependent processes. The mSDGM technique provides alternative information besides classical uni- and bivariate statistics and traditional multivariate dimensionality reduction techniques. 

The examples presented in the paper have been taken from forestry and several interesting conditional  structures have been detected. These findings could provide new theoretical and methodological ideas to the multivariate analysis of tree species for ecology, forestry and also spatial statistics. For both data sets, the SGDM has detected several interesting structures which should be tested by experts and could be interesting topics for future research. Both, the SDGM and mSDGM will be extended to alternative processes and thus provide a general formalism within spatial data analysis.

\section*{Acknowledgements}
We are grateful to Prof. Jim Clark at the Environmental Science Department in Duke University for providing us the Duke Forest data. We also extend our thanks to Prof. Alan Gelfand and Shinichiro Shirota helping us to obtain the data.   

\bibliographystyle{Chicago}
\bibliography{spatgraph}

\begin{thebibliography}{}

\bibitem[\protect\citeauthoryear{Aalen, Borgan, and Gjessing}{Aalen
  et~al.}{2008}]{Aalen2008}
Aalen, O.~O., {\O}.~Borgan, and H.~K. Gjessing (2008).
\newblock {\em Survival and Event History Analysis: A Process Point of View}.
\newblock Springer.

\bibitem[\protect\citeauthoryear{Ang}{Ang}{2010}]{Ang2010}
Ang, W. (2010).
\newblock {\em Statistical Methodologies for Events in a Linear Network}.
\newblock Ph.\ D. thesis, The University of Western Australia.

\bibitem[\protect\citeauthoryear{Ang, Baddeley, and Nair}{Ang
  et~al.}{2012}]{Ang2012}
Ang, W., A.~Baddeley, and G.~Nair (2012).
\newblock Geometrically corrected second order analysis of events on a linear
  network, with applications to ecology and criminology.
\newblock {\em Scandinavian Journal of Statistics\/}~{\em 39}, 591--617.

\bibitem[\protect\citeauthoryear{Baddeley}{Baddeley}{2010}]{Baddeley2010}
Baddeley, A. (2010).
\newblock {\em Handbook of Spatial Statistics}, Chapter Multivariate and Marked
  Point Processes, pp.\  371--402.
\newblock Chapman \& Hall/CRC Handbooks of Modern Statistical Methods. CRC
  Press.

\bibitem[\protect\citeauthoryear{Baddeley, Jammalamadaka, and Nair}{Baddeley
  et~al.}{2014}]{Baddeley2014}
Baddeley, A., A.~Jammalamadaka, and G.~Nair (2014).
\newblock Multitype point process analysis of spines on the dendrite network of
  a neuron.
\newblock {\em Journal of the Royal Statistical Society: Series C (Applied
  Statistics)\/}~{\em 63\/}(5), 673--694.

\bibitem[\protect\citeauthoryear{Baddeley, Rubak, and Turner}{Baddeley
  et~al.}{2015}]{Baddeley2015}
Baddeley, A., E.~Rubak, and R.~Turner (2015).
\newblock {\em Spatial Point Patterns: Methodology and Applications with R}.
\newblock CRC Press.

\bibitem[\protect\citeauthoryear{Banerjee, Carlin, and Gelfand}{Banerjee
  et~al.}{2004}]{Banerjee2004}
Banerjee, S., B.~P. Carlin, and A.~E. Gelfand (2004).
\newblock {\em Hierarchical Modeling and Analysis for Spatial Data}.
\newblock Boca Raton, London: Chapman \& Hall/CRC.

\bibitem[\protect\citeauthoryear{Bartlett}{Bartlett}{1963}]{Bartlett1963}
Bartlett, M.~S. (1963).
\newblock The spectral analysis of point processes.
\newblock {\em Journal of the Royal Statistical Society Series B\/}~{\em 29},
  264--296.

\bibitem[\protect\citeauthoryear{Bartlett}{Bartlett}{1964}]{Bartlett1964}
Bartlett, M.~S. (1964).
\newblock The spectral analysis of two-dimensional point processes.
\newblock {\em Biometrika\/}~{\em 51}, 299--311.

\bibitem[\protect\citeauthoryear{Bondy and Murty}{Bondy and
  Murty}{2008}]{Bondy2008}
Bondy, J.~A. and U.~S.~R. Murty (2008).
\newblock {\em Graph Theory}.
\newblock [New York, NY]: New York: Springer.

\bibitem[\protect\citeauthoryear{Brillinger}{Brillinger}{1972}]{Brillinger1972}
Brillinger, D. (1972).
\newblock The spectral analysis of stationary interval functions.
\newblock In {\em Proceedings of the Sixth Berkley Symposium}, Volume~1, pp.\
  483--513.

\bibitem[\protect\citeauthoryear{Brillinger}{Brillinger}{1981}]{Brillinger1981}
Brillinger, D. (1981).
\newblock {\em Time Series: Data Analysis and Theory}.
\newblock Holt, Rinchart and Winston, New York.

\bibitem[\protect\citeauthoryear{Brillinger}{Brillinger}{1996}]{Brillinger1996}
Brillinger, D. (1996).
\newblock Remarks concerning graphical models for time series and point
  processes.
\newblock {\em Revista de Econometrica\/}~{\em 16}, 1--23.

\bibitem[\protect\citeauthoryear{Brockwell and Davis}{Brockwell and
  Davis}{2006}]{Brockwell2006}
Brockwell, P.~J. and R.~A. Davis (2006).
\newblock {\em Time Series: Theory and Methods\/} (2nd ed.).
\newblock Springer.

\bibitem[\protect\citeauthoryear{Capobianco and Renshaw}{Capobianco and
  Renshaw}{1998}]{Capobianco1998}
Capobianco, R. and E.~Renshaw (1998).
\newblock The autocovariance function for marked point processes: A comparison
  between two different approaches.
\newblock {\em Biometrical Journal\/}~{\em 40\/}(4), 431--446.

\bibitem[\protect\citeauthoryear{Chatfield}{Chatfield}{1989}]{Chatfield1989}
Chatfield, C. (1989).
\newblock {\em The Analysis of Time Series: An Introduction}.
\newblock Chapman \& Hall CRC, Boca Raton.

\bibitem[\protect\citeauthoryear{Cowell, Dawid, Lauritzen, and
  Spiegelhalter}{Cowell et~al.}{1999}]{Cowell1999}
Cowell, R.~G., A.~P. Dawid, S.~L. Lauritzen, and D.~J. Spiegelhalter (1999).
\newblock {\em Probabilistic Networks and Expert Systems}.
\newblock Statistics for Engineering and Information Science. New York:
  Springer.

\bibitem[\protect\citeauthoryear{Cox and Wermuth}{Cox and
  Wermuth}{1996}]{Cox1996}
Cox, D. and N.~Wermuth (1996).
\newblock {\em Multivariate Dependencies : Models, Analysis and
  Interpretation}.
\newblock Chapman \& Hall CRC, Boca Raton.

\bibitem[\protect\citeauthoryear{Dahlhaus}{Dahlhaus}{2000}]{Dahlhaus2000}
Dahlhaus, R. (2000).
\newblock Graphical interaction models for multivariate time series.
\newblock {\em Metrika\/}~{\em 51\/}(2), 157--172.

\bibitem[\protect\citeauthoryear{Dahlhaus, Eichler, and
  Sandk{\"u}hler}{Dahlhaus et~al.}{1997}]{Dahlhausetal1997}
Dahlhaus, R., M.~Eichler, and J.~Sandk{\"u}hler (1997).
\newblock Identification of synaptic connections in neural ensembles by
  graphical models.
\newblock {\em Journal of Neuroscience Methods\/}~{\em 77\/}(1), 93--107.

\bibitem[\protect\citeauthoryear{Didelez}{Didelez}{2000}]{Didelez2000}
Didelez, V. (2000).
\newblock {\em Graphical Models for Event History Analysis Based on Local
  Independence}.
\newblock Ph.\ D. thesis, Universität Dortmund.

\bibitem[\protect\citeauthoryear{Didelez}{Didelez}{2007}]{Didelez2007}
Didelez, V. (2007).
\newblock Graphical models for composable finite markov processes.
\newblock {\em Scandinavian Journal of Statistics\/}~{\em 34}, 169--185.

\bibitem[\protect\citeauthoryear{Didelez}{Didelez}{2008}]{Didelez2008}
Didelez, V. (2008).
\newblock Graphical models for marked point processes based on local
  independence.
\newblock {\em Journal of the Royal Statistical Society Series B\/}~{\em 70},
  245--264.

\bibitem[\protect\citeauthoryear{Diestel}{Diestel}{2010}]{Diestel2010}
Diestel, R. (2010).
\newblock {\em Graph Theory.\/} (4. ed. ed.).
\newblock Heidelberg: Springer.

\bibitem[\protect\citeauthoryear{Diggle}{Diggle}{2002}]{Diggle2002}
Diggle, P. (2002).
\newblock {\em Statistical analysis spatial point patterns}.
\newblock Edward Arnold, London.

\bibitem[\protect\citeauthoryear{Diggle}{Diggle}{2013}]{Diggle2013}
Diggle, P. (2013).
\newblock {\em Statistical Analysis of Spatial and Spatio-Temporal Point
  Patterns}.
\newblock Chapman and Hall/CRC, Boca Raton.

\bibitem[\protect\citeauthoryear{Diggle, Zheng, and Durr}{Diggle
  et~al.}{2005}]{Diggle2005}
Diggle, P., P.~Zheng, and P.~Durr (2005).
\newblock Nonparametric estimation of spatial segregation in a multivariate
  point process: bovine tuberculosis in cornwall, uk.
\newblock {\em Journal of the Royal Statistical Society, Series C (Applied
  Statistics)\/}~{\em 54}, 645--658.

\bibitem[\protect\citeauthoryear{Dreassi and Gottard}{Dreassi and
  Gottard}{2007}]{Dreassi2007}
Dreassi, E. and A.~Gottard (2007).
\newblock A bayesian approach to model interdependent event histories by
  graphical models.
\newblock {\em Statistical Methods and Applications\/}~{\em 16\/}(1), 39--49.

\bibitem[\protect\citeauthoryear{Eckardt}{Eckardt}{2016a}]{Eckardt2016a}
Eckardt, M. (2016a).
\newblock {Graphical modelling of multivariate spatial point processes}.
\newblock {\em ArXiv e-prints\/}.

\bibitem[\protect\citeauthoryear{Eckardt}{Eckardt}{2016b}]{Eckardt2016}
Eckardt, M. (2016b).
\newblock {\em Reviewing Graphical Modelling of Multivariate Temporal
  Processes}, pp.\  221--229.
\newblock Cham: Springer International Publishing.

\bibitem[\protect\citeauthoryear{Eckardt}{Eckardt}{2016c}]{Eckardt2016d}
Eckardt, M. (2016c).
\newblock {\em sppgraph: Graphical modelling of multivariate spatial point
  patterns}.
\newblock R package version 1.0.

\bibitem[\protect\citeauthoryear{Eckardt and Mateu}{Eckardt and
  Mateu}{2016a}]{Eckardt2016b}
Eckardt, M. and J.~Mateu (2016a, July).
\newblock {Point patterns occurring on complex structures in space and
  space-time: An alternative network approach}.
\newblock {\em ArXiv e-prints\/}.

\bibitem[\protect\citeauthoryear{Eckardt and Mateu}{Eckardt and
  Mateu}{2016b}]{Eckardt2016c}
Eckardt, M. and J.~Mateu (2016b, July).
\newblock {Structured network regression for spatial point patterns}.
\newblock {\em ArXiv e-prints\/}.

\bibitem[\protect\citeauthoryear{Edwards}{Edwards}{2000}]{Edwards2000}
Edwards, D. (2000).
\newblock {\em Introduction to Graphical Modelling}.
\newblock Springer.

\bibitem[\protect\citeauthoryear{Eichler, Dahlhaus, and Sandk\"{u}hler}{Eichler
  et~al.}{2003}]{Eichler2003}
Eichler, M., R.~Dahlhaus, and J.~Sandk\"{u}hler (2003).
\newblock Partial correlation analysis for the identification of synaptic
  connections.
\newblock {\em Biological Cybernetics\/}~{\em 89}, 289--302.

\bibitem[\protect\citeauthoryear{Fosen, Borgan, Weedon-Fekj{\ae}r, and
  Aalen}{Fosen et~al.}{2006}]{Fosen2006}
Fosen, J., {\O}.~Borgan, H.~Weedon-Fekj{\ae}r, and O.~O. Aalen (2006).
\newblock Dynamic analysis of recurrent event data using the additive hazard
  model.
\newblock {\em Biometrical Journal\/}~{\em 48\/}(3), 381--398.

\bibitem[\protect\citeauthoryear{Gottard}{Gottard}{2007}]{Gottard2007}
Gottard, A. (2007).
\newblock On the inclusion of bivariate marked point processes in graphical
  models.
\newblock {\em Metrika\/}~{\em 66\/}(3), 269--287.

\bibitem[\protect\citeauthoryear{Gottard and Rampichini}{Gottard and
  Rampichini}{2007}]{Gottard2007a}
Gottard, A. and C.~Rampichini (2007).
\newblock Chain graphs for multilevel models.
\newblock {\em Statistics \& Probability Letters\/}~{\em 77\/}(3), 312 -- 318.

\bibitem[\protect\citeauthoryear{Grabarnik and S\"{a}rkk\"{a}}{Grabarnik and
  S\"{a}rkk\"{a}}{2009}]{Grabarnik2009}
Grabarnik, P. and A.~S\"{a}rkk\"{a} (2009).
\newblock Modelling the spatial structure of forest stands by multivariate
  point processes with hierarchical interactions.
\newblock {\em Ecological Modelling\/}~{\em 220}, 1232--1240.

\bibitem[\protect\citeauthoryear{Guan}{Guan}{2006}]{Guan2006}
Guan, Y. (2006).
\newblock Tests for independence between marks and points of a marked point
  process.
\newblock {\em Biometrics\/}~{\em 62\/}(1), 126--134.

\bibitem[\protect\citeauthoryear{Guan and Afshartous}{Guan and
  Afshartous}{2007}]{Guan2007a}
Guan, Y. and D.~R. Afshartous (2007).
\newblock Test for independence between marks and points of marked point
  processes: a subsampling approach.
\newblock {\em Environmental and Ecological Statistics\/}~{\em 14}, 101--111.

\bibitem[\protect\citeauthoryear{Illian and Burslem}{Illian and
  Burslem}{2007}]{Illian2007}
Illian, J. and D.~Burslem (2007).
\newblock Contributions of spatial point process modelling to biodiversity
  theory.
\newblock {\em Journal de la Soci{\'e}t{\'e} Fran\c{c}aise de
  Statistique\/}~{\em 148}, 9--29.

\bibitem[\protect\citeauthoryear{Illian, Penttinen, Stoyan, and Stoyan}{Illian
  et~al.}{2008}]{Illian2008}
Illian, J., A.~Penttinen, H.~Stoyan, and D.~Stoyan (2008).
\newblock {\em Statistical Analysis and Modelling of Spatial Point Patterns}.
\newblock John Wiley \& Sons, New York.

\bibitem[\protect\citeauthoryear{Koller and Friedman}{Koller and
  Friedman}{2010}]{Koller2010}
Koller, D. and N.~Friedman (2010).
\newblock {\em Probabilistic Graphical Models: Principles and Techniques}.
\newblock MIT Press.

\bibitem[\protect\citeauthoryear{Lauritzen}{Lauritzen}{1996}]{Lauritzen1996}
Lauritzen, S.~L. (1996).
\newblock {\em Graphical Models}.
\newblock Oxford University Press.

\bibitem[\protect\citeauthoryear{Leininger}{Leininger}{2014}]{Leininger2014}
Leininger, T.~J. (2014).
\newblock {\em Bayesian Analysis of Spatial Point Patterns}.
\newblock Ph.\ D. thesis, Duke University.

\bibitem[\protect\citeauthoryear{Leininger and Gelfand}{Leininger and
  Gelfand}{2016}]{Leininger2016}
Leininger, T.~J. and A.~E. Gelfand (2016).
\newblock Bayesian inference and model assessment for spatial point patterns
  using posterior predictive samples.
\newblock {\em Bayesian Analysis\/}, Advance Publication.

\bibitem[\protect\citeauthoryear{Marchette}{Marchette}{2004}]{Marchette2004}
Marchette, D. (2004).
\newblock {\em Random Graphs for Statistical Pattern Recognition}.
\newblock J. Wiley \& Sons, Hoboken, NJ.

\bibitem[\protect\citeauthoryear{Martinussen}{Martinussen}{2010}]{Martinussen2010}
Martinussen, T. (2010).
\newblock Dynamic path analysis for event time data: large sample properties
  and inference.
\newblock {\em Lifetime Data Analysis\/}~{\em 16\/}(1), 85--101.

\bibitem[\protect\citeauthoryear{Mateu}{Mateu}{2000}]{Mateu2000}
Mateu, J. (2000).
\newblock Second-order characteristics of spatial marked processes with
  applications.
\newblock {\em Nonlinear Analysis: Real World Applications\/}~{\em 1},
  145--162.

\bibitem[\protect\citeauthoryear{Møller, Ghorbani, and Rubak}{Møller
  et~al.}{2016}]{MollerGhorbaniRubak2016}
Møller, J., M.~Ghorbani, and E.~Rubak (2016).
\newblock Mechanistic spatio-temporal point process models for marked point
  processes, with a view to forest stand data.
\newblock {\em Biometrics\/}~{\em 72\/}(3), 687--696.

\bibitem[\protect\citeauthoryear{M{\o}ller and Waagepetersen}{M{\o}ller and
  Waagepetersen}{2004}]{Moller2004}
M{\o}ller, J. and R.~P. Waagepetersen (2004).
\newblock {\em Statistical Inference and Simulation for Spatial Point
  Processes}.
\newblock Chapman and Hall/CRC, Boca Raton.

\bibitem[\protect\citeauthoryear{Mugglestone and Renshaw}{Mugglestone and
  Renshaw}{2001}]{Mugglestone2001}
Mugglestone, M. and E.~Renshaw (2001).
\newblock Spectral tests of randomness for spatial point pattern.
\newblock {\em Environmental and Ecological Statistics\/}~{\em 8}, 237--251.

\bibitem[\protect\citeauthoryear{Mugglestone and Renshaw}{Mugglestone and
  Renshaw}{1996a}]{Mugglestone1996a}
Mugglestone, M.~A. and E.~Renshaw (1996a).
\newblock The exploratory analysis of bivariate spatial point pattern using
  cross-spectra.
\newblock {\em Environmetrics\/}~{\em 7}, 361--377.

\bibitem[\protect\citeauthoryear{Mugglestone and Renshaw}{Mugglestone and
  Renshaw}{1996b}]{Mugglestone1996b}
Mugglestone, M.~A. and E.~Renshaw (1996b).
\newblock A practical guide to the spectral analysis of spatial point
  processes.
\newblock {\em Computational Statistics and Data Analysis\/}~{\em 21}, 43--65.

\bibitem[\protect\citeauthoryear{Myllymäki}{Myllymäki}{2009}]{Myllymaeki2009}
Myllymäki, M. (2009).
\newblock {\em Statistical Models and Inference for Spatial Point Patterns with
  Intensity-Dependent Marks}.
\newblock Ph.\ D. thesis, University of Jyväskylä.

\bibitem[\protect\citeauthoryear{Okabe and Yamada}{Okabe and
  Yamada}{2001}]{Okabe2001}
Okabe, A. and I.~Yamada (2001).
\newblock The ${K}$-function on a network and its computational implementation.
\newblock {\em Geographical Analysis\/}~{\em 33\/}(3), 271--290.

\bibitem[\protect\citeauthoryear{Palmer}{Palmer}{1990}]{Palmer1990}
Palmer, M.~W. (1990).
\newblock Vascular flora of the duke forest, north carolina.
\newblock {\em Castanea\/}~{\em 55\/}(4), 229--244.

\bibitem[\protect\citeauthoryear{Pearl}{Pearl}{1988}]{Pearl1988}
Pearl, J. (1988).
\newblock {\em Probabilistic Reasoning in Intelligent Systems: Networks of
  Plausible Inference}.
\newblock Morgan Kaufmann Publishers Inc.

\bibitem[\protect\citeauthoryear{Penrose}{Penrose}{2003}]{Penrose2003}
Penrose, M. (2003).
\newblock {\em Random geometric graphs.}
\newblock Oxford University Press, Oxford.

\bibitem[\protect\citeauthoryear{Penrose}{Penrose}{2005}]{Penrose2005}
Penrose, M. (2005).
\newblock Multivariate spatial central limit theorems with application to
  percolation and spatial graphs.
\newblock {\em Annals of Probability\/}~{\em 33}, 1945--1991.

\bibitem[\protect\citeauthoryear{Penrose and Yukich}{Penrose and
  Yukich}{2001}]{Penrose2001}
Penrose, M.~D. and J.~E. Yukich (2001).
\newblock Central limit theorems for some graph in computational geometry.
\newblock {\em Annals of Applied Probability\/}~{\em 11}, 1005--1041.

\bibitem[\protect\citeauthoryear{Penttinen, Stoyan, and Henttonen}{Penttinen
  et~al.}{1992}]{Penttinen1992}
Penttinen, A., D.~Stoyan, and H.~M. Henttonen (1992).
\newblock Marked point processes in forest statistics.
\newblock {\em Forest Science\/}~{\em 38}, 806--824.

\bibitem[\protect\citeauthoryear{Priestley}{Priestley}{1981}]{Priestley1981}
Priestley, M. (1981).
\newblock {\em Spectral Analysis and Time Series}.
\newblock Academic Press, London.

\bibitem[\protect\citeauthoryear{Renshaw}{Renshaw}{1997}]{Renshaw1997}
Renshaw, E. (1997).
\newblock Spectral techniques in spatial analysis.
\newblock {\em Forest Ecology ad Management\/}~{\em 94}, 165--174.

\bibitem[\protect\citeauthoryear{Renshaw}{Renshaw}{2002}]{Renshaw2002}
Renshaw, E. (2002).
\newblock Two-dimensional spectral analysis for marked point processes.
\newblock {\em Biometrical Journal\/}~{\em 44}, 718--745.

\bibitem[\protect\citeauthoryear{Renshaw and Ford}{Renshaw and
  Ford}{1984}]{Renshaw1984}
Renshaw, E. and E.~Ford (1984).
\newblock The description of spatial pattern using two-dimensional spectral
  analysis.
\newblock {\em Vegetatio\/}~{\em 56}, 75--85.

\bibitem[\protect\citeauthoryear{Renshaw and Ford}{Renshaw and
  Ford}{1983}]{Renshaw1983}
Renshaw, E. and E.~D. Ford (1983).
\newblock The interpretation of process from pattern using two-dimensional
  spectral analysis: {M}ethods and problems of interpretation.
\newblock {\em Applied Statistics\/}~{\em 32}, 51--63.

\bibitem[\protect\citeauthoryear{Ripley}{Ripley}{1976}]{Ripley1976}
Ripley, B.~D. (1976).
\newblock The second-order analysis of stationary point processes.
\newblock {\em Journal of Applied Probability\/}~{\em 13}, 255--266.

\bibitem[\protect\citeauthoryear{Saura and Mateu}{Saura and
  Mateu}{2006}]{Saura2006}
Saura, F. and J.~Mateu (2006).
\newblock Estimating mark functions through spectral analysis for marked point
  patterns.
\newblock {\em Communications in Statistics - Theory and Methods\/}~{\em
  35\/}(5), 861--885.

\bibitem[\protect\citeauthoryear{Schlather}{Schlather}{2001}]{Schlather2001}
Schlather, M. (2001, 02).
\newblock On the second-order characteristics of marked point processes.
\newblock {\em Bernoulli\/}~{\em 7\/}(1), 99--117.

\bibitem[\protect\citeauthoryear{Schlather, Riberio, and Diggle}{Schlather
  et~al.}{2004}]{Schlather2004}
Schlather, M., P.~Riberio, and P.~Diggle (2004).
\newblock Detecting dependence between marks and locations of marked point
  processes.
\newblock {\em Journal of the Royal Statistical Society, series B\/}~{\em 66},
  79--93.

\bibitem[\protect\citeauthoryear{{Shirota} and {Gelfand}}{{Shirota} and
  {Gelfand}}{2016}]{Shirota2016}
{Shirota}, S. and A.~E. {Gelfand} (2016, June).
\newblock {Approximate Marginal Posterior for Log Gaussian Cox Processes}.
\newblock {\em ArXiv e-prints\/}.

\bibitem[\protect\citeauthoryear{Spirtes}{Spirtes}{2000}]{Spirtes2000}
Spirtes, P. (2000).
\newblock {\em Causation, Prediction, and Search\/} (2. ed. ed.).
\newblock Cambridge. Mass.: Cambridge. Mass.: MIT Press.

\bibitem[\protect\citeauthoryear{Stoyan, Kendall, and Mecke}{Stoyan
  et~al.}{1995}]{Stoyan1995}
Stoyan, D., W.~S. Kendall, and J.~Mecke (1995).
\newblock {\em Stochastic Geometry and Its Applications\/} (Second ed.).
\newblock Wiley, Chichester.

\bibitem[\protect\citeauthoryear{Stoyan and Stoyan}{Stoyan and
  Stoyan}{1994}]{StoyanStoyan1994}
Stoyan, D. and H.~Stoyan (1994).
\newblock {\em Fractals, Random Shapes, and Point Fields : Methods of
  Geometrical Statistics}.
\newblock Chichester, New York: Wiley.

\bibitem[\protect\citeauthoryear{Stoyan and W\"{a}lder}{Stoyan and
  W\"{a}lder}{2000}]{Stoyan2000}
Stoyan, D. and O.~W\"{a}lder (2000).
\newblock On variograms in point process statistics, ii: Models for markings
  and ecological interpretation.
\newblock {\em Biometrical Journal\/}~{\em 42}, 171--187.

\bibitem[\protect\citeauthoryear{Terres and Gelfand}{Terres and
  Gelfand}{2016}]{Terres2016}
Terres, M.~A. and A.~E. Gelfand (2016).
\newblock Spatial process gradients and their use in sensitivity analysis for
  environmental processes.
\newblock {\em Journal of Statistical Planning and Inference\/}~{\em 168}, 106
  -- 119.

\bibitem[\protect\citeauthoryear{Whittaker}{Whittaker}{2008}]{Whittaker2008}
Whittaker, J.~C. (2008).
\newblock {\em Grapical Models in Applied Multivariate Statistics}.
\newblock John Wiley \& Sons.

\bibitem[\protect\citeauthoryear{Wiegand, Gunatilleke, and Gunatilleke}{Wiegand
  et~al.}{2007}]{Wiegand2007}
Wiegand, T., S.~Gunatilleke, and N.~Gunatilleke (2007).
\newblock Species associations in a heterogeneous sri lankan dipterocarp
  forest.
\newblock {\em The American Naturalist\/}~{\em 170\/}(4), E77--E95.

\bibitem[\protect\citeauthoryear{Xi, Peet, and Urban}{Xi et~al.}{2008}]{Xi2008}
Xi, W., R.~K. Peet, and D.~L. Urban (2008).
\newblock Changes in forest structure, species diversity and spatial pattern
  following hurricane disturbance in a piedmont north carolina forest, usa.
\newblock {\em Journal of Plant Ecology\/}~{\em 1\/}(1), 43--57.

\end{thebibliography}

\end{document}